\documentclass[preprint,endfloats,showpacs,eqsecnum,pre,draft,intlimits,amsmath,amssymb]{revtex4}
\usepackage{bm}
\usepackage[final]{graphicx}
\usepackage{longtable}
\newcommand{\be}{\begin{equation}}
\newcommand{\ee}{\end{equation}}
\newcommand{\bea}{\begin{eqnarray}}
\newcommand{\eea}{\end{eqnarray}}

\newcommand{\eqan}[1]{\begin{eqnarray*}#1\end{eqnarray}}

\newcommand{\eq}[1]{\begin{equation}#1\end{equation}}

\begin{document}

\title{Universal amplitude in density-force relations for polymer chains in confined geometries: Massive field theory approach.}

\author{Z.Usatenko}
 \affiliation{Institute for Condensed Matter
Physics, National Academy of Sciences of Ukraine, 79011 Lviv,
Ukraine}

\vspace{0.1cm}
\date{\today}

\begin{abstract}

The universal density-force relation is analyzed and the
correspondent universal amplitude ratio $B_{real}$ is obtained using
the massive field theory approach in fixed space dimensions $d=3$ up
to one-loop order. The layer monomer density profiles of ideal
chains and real polymer chains with excluded volume interaction in a
good solvent between two parallel repulsive walls, one repulsive and
one inert wall are obtained. Besides, taking into account the
Derjaguin approximation the layer monomer density profiles for
dilute polymer solution confined in semi-infinite space containing
mesoscopic spherical particle of big radius are calculated. The last
mentioned situation is analyzed for both cases when wall and
particle are repulsive and for the mixed case of repulsive wall and
inert particle. The obtained results are in good agreement with
previous theoretical results and with the results of Monte Carlo
simulations.

\end{abstract}
\vspace{0.2cm} 

\maketitle
\renewcommand{\theequation}{\arabic{equation}}
The investigation of polymer solution near the surfaces and in film
geometries or mesoscopic particles dissolved in the solution is a
task of great interest from theoretical and application point of
view.

The monomer density profiles of dilute polymer solution bounded by a
planar repulsive wall has a depletion region of mesoscopic width of
order of the coil size $R_{x}\approx N^{\nu}$ (where $N$ is the
number of monomers per chain), and for the distances $\tilde{z}$
from the wall that are small compared to this width but bigger than
microscopic lengths of monomer size $\tilde{l}$  the profile
increases as
$$\rho(\tilde{z})\sim \tilde{z}^{\frac{1}{\nu}}$$ with Flory exponent $\nu$
($\nu$ is $1/2$ for ideal polymer chains and $\nu\approx0.588$ for
polymer chains with excluded volume interaction (EVI)).
 This remarkable theoretical predictions was proposed by
Joanny, Leibler and de Gennes \cite{JLG}. They also mentioned that
the monomer density close to the wall is proportional to the force
per unit area which the polymer solution exerts on the wall. But,
for the first time a complete quantitative expression for the
universal density-force relation was obtained by Eisenriegler on the
basis of $\epsilon$- expansion up to first order in \cite{E97}. As
was mentioned in  \cite{E97}, the correspondent density-force
relations with the same universal amplitude $B$ are valid for the
different cases: 1) a single polymer chain with one end (or both
ends) fixed in the half space bounded by the wall; 2) a single chain
trapped in the slit of two walls; 3) for the case of dilute and
semi-dilute solution of free polymer chains in a half space; 4) for
the case of polymer chain in a half space containing a mesoscopic
particle of arbitrary shape. The verification of the universal
density-force relation was performed by simulation techniques using
an off-lattice bead-spring model of a polymer chain trapped between
two parallel repulsive walls \cite{MB98} and by the lattice Monte
Carlo algorithm on a regular cubic lattice in three dimensions
\cite{HG04}. Unfortunately, the obtained results of Monte Carlo
simulations in \cite{MB98} are much higher $B_{MC}>2.48$ than
theoretical predictions $B_{\epsilon}\approx 1.85$ obtained in
\cite{E97}. As it was mentioned in \cite{MB98}, there is a
systematic decrease of $B_{eff}$ with increasing the distance $L$
between the walls. In accordance with it some rough linear
extrapolation with plotting $B_{eff}$ versus $1/\sqrt{L}$ which
yields an extrapolated result $B\sim 1.4$ was performed in
\cite{MB98}. Recent numerical results based on the lattice Monte
Carlo algorithm on a regular cubic lattice gives the value $B\approx
1.70\pm 0.08$ which is smaller than previous theoretical
predictions. This all indicates that the above mentioned task still
have a lot of open questions and the present paper tries to give
answers for some of them.

The present paper is devoted to investigation of the universal
density-force relation and calculation of the universal amplitude
$B$ in the framework of the massive field theory approach. The
massive field theory approach gives better agreement with the
experimental data and the results of Monte Carlo calculations as it
was shown in the case of infinite \cite{Par80,Parisi}, semi-infinite
\cite{DSh98} systems, and specially in the case of dilute polymer
solutions in semi-infinite geometry \cite{U06} and confined geometry
\cite{DU09}. As was mentioned above, the knowledge of the universal
amplitude $B$ allows to obtain the monomer density profiles for
whole class of different systems (cases 1)-4)). Besides, it should
be mentioned that the universal density-force relation is valid not
only for the case of polymer chains trapped in between two repulsive
walls but also for the mixed case of one repulsive and one inert
wall.

We consider a dilute polymer solution, where different polymer
chains do not overlap and the behavior of such polymer solution can
be described by a single polymer chain. As it is known, the single
polymer chain can be modeled by the model of random walk and this
corresponds to the ideal chain at $\theta$-solvent or self-avoiding
walk for the real polymer chain with EVI for temperatures above the
$\theta$-point. Taking into account the polymer-magnet analogy
developed by \cite{deGennes}, their scaling properties in the limit
of an infinite number of steps $N$ may be derived by a formal $n \to
0$ limit of the field theoretical $\phi^4$ $O(n)$- vector model at
its critical point. It should be mentioned that $1/N$ plays the role
of a critical parameter  analogous to the reduced  critical
temperature in magnetic systems. The role of a second critical
parameter plays the deviation from the adsorption
 threshold $(c\propto(T-T_a)/T_a)$ (where $T_{a}$ is adsorption temperature). The value $c$
corresponds to the adsorption energy
 divided by $k_{B}T$ (or the surface enhancement constant in field theoretical
 treatment). The adsorption threshold for long-flexible
 polymer chains, where $1/N\to 0$ and $c\to 0$ is a multicritical phenomenon.

 In order to obtain the
universal amplitude $B$ in  universal monomer-density relation let's
consider the single polymer chain with one end fixed fluctuating
near the repulsive wall such that
$(\tilde{l}<<\tilde{z}<<z<<R_{x})$.

The correspondent layer monomer densities
$\rho_{\lambda}(\tilde{z})$ defined by \cite{E97} is:
 \be \rho_{\lambda}(\tilde{z})d \tilde{z}=\frac{(R_{x})^{1/\nu}}{N}d
N_{\lambda}(\tilde{z})\label{md_defin},\ee

where $d N_{\lambda}(\tilde{z})$ is the number of monomers in the
layer between $\tilde{z}$ and $\tilde{z}+d\tilde{z}$, $R_{x}$ is the
projection of the end to end distance ${\bf R}$ onto the direction
of $x$ axis. Besides, $\rho_{\lambda}$ is obtained from monomer
density $\rho(\tilde{{\bf r}},\tilde{z})$ after integration over the
$d-1$ components parallel to the wall. The scaling dimension of
$\rho(\tilde{{\bf r}},\tilde{z})$ is $l^{1/\nu-d}$ and equals the
ordinary dimensions of the quantity \be{\Psi(\tilde{{\bf
x}})}=\frac{R_{x}^{1/\nu}}{2L_{0}}{\bf \Phi}^2 (\tilde{{\bf x}}),
\ee

where $L_{0}$ is the conjugate Laplace variable which has the
dimension of length squared and is proportional to the total number
of monomers $N$ of the polymer chain.
 Following the description of the problem as given in
\cite{E97}, the monomer density in this case is \be
<\rho(\tilde{{\bf x}})>=\frac{{\cal IL}<\Psi(\tilde{{\bf
x}})\cdot{\vec \phi}({\bf x}){\vec\phi}({\bf x'})>_{w}}{{\cal
IL}<{\vec \phi}({\bf x}){\vec\phi}({\bf x'})>_{w}}\label{mon_dens}
\ee in the limit $n\to 0$. The average $<{\cdot}>_{w}$ in
(\ref{mon_dens}) denotes a statistical average for a Ginzburg-Landau
field in semi-infinite geometry. The dot in (\ref{mon_dens}) means a
cumulant average. The correspondent Ginzburg-Landau Hamiltonian
describing the system in semi-infinite ($j=1$) or confined geometry
of two walls ($j=1,2$) is: \bea {\cal H}[{\vec \phi}] &=&\int
d^{d}x\bigg\lbrace \frac{1}{2} \left( \nabla{\vec{\phi}} \right)^{2}
+\frac{{\mu_{0}}^{2}}{2} {\vec{\phi}}^{2} +\frac{v_{0}}{4!}
\left({\vec{\phi}}^2 \right)^{2}
\bigg\rbrace \nonumber\\
&+&\sum_{j=1}^{2}\frac{c_{j_{0}}}{2} \int d^{d-1}r {\vec{\phi}}^{2},
\label{hamiltonian}\eea

where ${\vec \phi}({\bf x})$ is an $n$-vector field with the
components $\phi_i(x)$, $i=1,...,n$ and  ${\bf{x}}=({\bf r},z)$,
$\mu_0$ is the "bare mass", $v_0$ is the bare coupling constant
which characterizes the strength of the excluded volume interaction
(EVI). The surfaces of the system is characterized by a certain
surface enhancement constant $c_{j_{0}}$, where $j=1,2$.
 The interaction between the polymer chain
and the walls is implemented by the different boundary conditions.

In the case of two repulsive walls (where the segment partition
function and thus the partition function for the whole polymer chain
 tends to 0 as any segment approaches the surface of the walls)
Dirichlet-Dirichlet boundary conditions (D-D b.c.) takes place:

\be c_{1}\to +{\infty},\quad c_{2}\to +{\infty}\quad or\quad{\vec
\phi}({\bf {r}},0)={\vec \phi}({\bf {r}},L)=0\label{DD}, \ee
 and for the mixed case of one repulsive and one
inert wall Dirichlet-Neumann boundary conditions(D-N b.c.) are: \be
c_{1}\to +{\infty},\quad c_{2}=0\quad or\quad {\vec \phi}({\bf
{r}},0)=0, \quad\frac{\partial{{\vec \phi}({\bf {r}},z)}}{\partial
z}|_{z=L}=0\label{DN}.\ee

Taking into account that $\int d^{d}\tilde{{\bf x}}<\rho(\tilde{{\bf
x}})>=R_{x}^{1/\nu}$ the property \be  \int d^{d}\tilde{{\bf x}}
{\cal IL} <\Psi(\tilde{{\bf x}})\cdot{\vec \phi}({\bf
x}){\vec\phi}({\bf x'})>_{w}=R_{x}^{1/\nu}{\cal{IL}}<{\vec
\phi}({\bf x}){\vec\phi}({\bf x'})>_{w},\label{mon_norm}\ee takes
place. It should be mentioned that near the repulsive wall the
short-distance expansion of ${\bf \Phi}^2$ takes place and it
assumes \cite{DD81,C90,E93,EKD96} \be \Psi(\tilde{{\bf
r}},\tilde{z})\to B \tilde{z}^{1/\nu} \frac{[{\bf
\Phi_{\perp}}(\tilde{{\bf r}})]^{2}}{2},\label{short_dist_exp}\ee
for the distances $\tilde{l}<<\tilde{z}$, where $\tilde{l}$ is
monomer size. The surface operator $\frac{[{\bf
\Phi_{\perp}}(\tilde{{\bf r}})]^{2}}{2}$ with ${\bf
\Phi_{\perp}}=\frac{\partial{\bf \Phi}(\tilde{{\bf
r}},\tilde{z})}{{\partial \tilde{z}}}|_{\tilde{z}=0}$ is the
component of the stress tensor perpendicular to the walls. Taking
into account the shift identity \cite{DDE83,D86,E97} for the case of
semi-infinite geometry \be \int d^{d-1}\tilde{{\bf r}} <\frac{[{\bf
\Phi_{\perp}}(\tilde{{\bf r}})]^{2}}{2} \cdot {\vec \phi}({\bf
x}){\vec\phi}({\bf x'})>_{w}=(\frac{\partial}{\partial
z}+\frac{\partial}{\partial z'})<{\vec \phi}({\bf x}){\vec\phi}({\bf
x'})>_{w} \ee and integrating it over $\int d^{d}{\bf x}'$ for the
layer monomer density Eq.(\ref{md_defin}) in accordance with
Eqs.(\ref{mon_dens}), (\ref{short_dist_exp}) the universal
density-force relation can be obtained \be
<\rho_{\lambda}(\tilde{z})>=B \tilde{z}^{1/\nu}
\frac{f}{k_{B}T},\label{mn_dens_rel}\ee where \be
\frac{f}{k_{B}T}=\frac{\partial}{\partial z} \ln [{\cal IL} \int
d^{d}{\bf x}'<{\vec \phi}({\bf x}){\vec\phi}({\bf
x'})>_{w}]\label{force}\ee is the force per area that the polymer
chains exert on the wall. It should be mentioned that the
density-force relation (\ref{mn_dens_rel}) takes place for the
distances ${\tilde l}<<\tilde{z}<<R_{x}$, and $B$ is universal
amplitude. Following the scheme of obtaining the universal amplitude
$B$ as it was proposed in \cite{D86,E93} the correspondent
calculations based on the massive field theory approach in fixed
$d=3$ dimensions were performed. Thus, for the layer monomer density
of ideal polymer chain takes place: \be
<\rho_{\lambda}(\tilde{z})>_{G}=\frac{<R_{x}^{2}>^{1/2\nu}}{L_{0}}\frac{{\cal
IL}G( ;z,\tilde{z})\int_{0}^{\infty}dz' G( ;\tilde{z},z')}{{\cal IL}
\int_{0}^{\infty}dz'G(
;z,z')}=2\frac{\tilde{z}^{2}}{z}.\label{mon_dens_ideal} \ee  Taking
into account the value of $f/k_{B}T\approx 1/z$ for the force
exerted by ideal polymer chain on the wall from
Eq.(\ref{mn_dens_rel}) the universal amplitude $B$ can be obtained:
\be B_{id}=2. \label{bideal}\ee

 The case of real polymer chain is more complicated, because EVI
with nonequal to zero the bare coupling constant $v_{0}$ should be
taken into account. Taking into account Eq.(\ref{mn_dens_rel}) and
Eqs.(\ref{mon_dens}),(\ref{force}) after renormalization of the mass
\be \mu_{0}^{2}=\mu^{2}-\frac{v_{0}}{3} J_{1}(\mu_{0})+O(v_{0}^2),
\ee where $J_{1}(\mu_{0})=\frac{1}{(2
\pi)^{d-1}}\int\frac{d^{d-1}q}{2\kappa_{q}}$ with $\kappa
_{q}=\sqrt{q^{2}+\mu_{0}^{2}}$, the renormalization of the coupling
constant $v_{0}=v\mu$ and including the correspondent UV-finite
renormalization factors (see \cite{U10}) in the limit $n\to 0$ up to
one-loop order for the real polymer chain with EVI the universal
amplitude $B_{real}$ can be obtained : \be B=B_{real}=
2(1-\frac{\tilde{v}}{4}(1+\frac{\ln{2}}{2}-\gamma_{E})),\label{bevif}\ee
where $\gamma_{E}= 0.577$ is Euler's constant. Here we took into
account that $\nu=\frac{1}{2}(1+\frac{\tilde{v}}{8})+O(v^{2})$ and
the following definition $v=b_{n}(d){\tilde{v}}$ was introduced with
$b_{n}(d)=\frac{6}{n+8}\frac{(4\pi)^{d/2}}{\Gamma[\epsilon/2]}$. The
correspondent fixed point is equal: ${\tilde{v}}^{*}=1$. At $d=3$
Eq.(\ref{bevif}) leads to: \be B_{real}\approx 1.62.\label{bevi}\ee

The obtained result Eq.({\ref{bevi}}) is in agreement with the
result obtained by  Eisenriegler \cite{E97} for real polymer chains
in the framework of $\epsilon$-expansion at $d=3$:
$B=B_{real}^{(\epsilon)}=2(1-b\epsilon)\approx 1.85$, where
$\epsilon=4-d$ and
 $b=(\ln 2+\gamma_{E}-2/3)/8$. As it is easy to see, the result
obtained in the framework of the massive field theory approach is
slightly smaller than result of $\epsilon$-expansion \cite{E93,E97}
and is in agreement with the numerical result $B_{real}\approx
1.70\pm 0.08$ of Monte Carlo simulations \cite{HG04} and with the
result of rough linea extrapolation $B_{extr}\sim 1.4$ obtained on
the basis of plotting $B_{eff}$ versus $1/\sqrt{L}$ in\cite{MB98}.

 As it was shown in \cite{E97},
the density-force relation (\ref{mn_dens_rel}) is also valid for the
case of dilute and monodisperse solution of free chains in
semi-infinite space. The polymer density far from the wall is fixed
and the pressure on the wall $f/A$ (where $A$ is the area of the
wall) equals the pressure in the bulk $n_{B}k_{B}T$. Thus, the
monomer density of dilute polymer solution of free chains in
semi-infinite space according to Eq.(\ref{mn_dens_rel}) is \be
<\rho(\tilde{{\bf x}})>_{f}=B
\tilde{z}^{1/\nu}n_{B},\label{mn_free}\ee where $n_{B}$ is the
polymer density in bulk far from the wall.

In the case of a spherical particle with radius $R$ much larger than
the distance of its closest point $a$ to the surface and much larger
than radius of gyration $R_{g}$ the Derjaguin approximation, which
describes the sphere by a superposition of fringes with local
distance from the wall
$L({\bf{r}}_{\parallel})=a+{\bf{r}}_{\parallel}^2/(2{\bf R})$,
should be applied \cite{D34}.

Taking into account the depletion interaction potential between the
particle and the wall which we obtained in \cite{DU09}(see
Eq.(7.12)):

\be \frac{\phi_{depl}(a)}{n_{B}k_{B}T}=2\pi R
R_{x}^{2}\int_{a/R_{x}}^{\infty} d y \Theta (y),\ee
 and the correspondent scaling function for the free energy of
interaction for the slit geometry $\Theta(y)$ (see Eqs.(5.6),(5.8)
and Eqs.(6.4),(6.8) in \cite{DU09}) the layer monomer density of
dilute polymer solution in semi-infinite space containing spherical
particle of big radius for $A=1$ and $n_{B}=1/(L A)$ can be
obtained: \be <\rho_{\lambda}(\tilde{z})>_{wp}=\frac{B
\tilde{z}^{1/\nu}}{L}(1-2\pi R
a^{2}\Theta(\frac{a}{R_{x}})).\label{mon_dens_part_f}\ee

As it is easy to see from Eq.(\ref{mon_dens_part_f}), the layer
monomer density depends not only on $R_{x}$, but also on the shape
of the mesoscopic particle and its distance from the wall.

The density-force relation in the case of single polymer chain
trapped in the slit of two walls can be shown in the same way,
because the Eq.(\ref{mon_norm}) and Eq.(\ref{short_dist_exp}) take
place not only for the averages of the type $<\cdot>_{w}$, but also
for the averages $<\cdot>_{ww}$. More detailed calculations can be
found in \cite{U10}.
 As it was mentioned in \cite{E97}, the
monomer density in the case of dilute polymer solution between two
repulsive walls has a maximum in the center $\tilde{z}=L/2$ of the
slit. If the distant wall at $\tilde{z}=L$ is inert or in another
words is at the adsorption threshold, the density-force relation
Eq.(\ref{mn_dens_rel}) again takes place with the same values of B
(see Eqs.({\ref{bideal}),(\ref{bevi})), as it was mentioned by
Eisenrieglier \cite{E97}. In this last mentioned case the polymer
chain prefers the distant inert wall and the monomer density maximum
is at $\tilde{z}=L$. The results of calculations of the layer
monomer density profiles for the case of ideal polymer chain and
real polymer chain with EVI confined between two repulsive walls
(D-D b.c), one repulsive and one inert wall (D-N b.c.) in accordance
with Eq.(\ref{mn_dens_rel}) and taking into account the
correspondent values of $B_{ideal}$ and $B_{real}$ (see
Eqs.(\ref{bideal}), (\ref{bevi})) are presented on Fig.1 and Fig.2,
respectively. Besides, Fig.1 and Fig.2 present results for the case
of ideal and real polymer chains in semi-infinite space containing
spherical particle of big radius. The last mentioned situation is
analyzed for both cases when wall and particle are repulsive and for
the mixed case of repulsive wall and inert particle.

The obtained results (see Fig.1) indicate that the layer monomer
density profiles for ideal polymer chains are weaker then for real
polymer chains with EVI in the case of two repulsive walls (or D-D
b.c.). Completely opposite behavior of monomer density profiles is
observed for the case of one repulsive and one inert wall (or D-N
b.c.), as it is easy to see from Fig.2. Besides, the layer monomer
densities for curvature surfaces are smaller then for planar
surfaces.
\begin{figure}[ht!]
\begin{center}
\includegraphics[width=8.0cm]{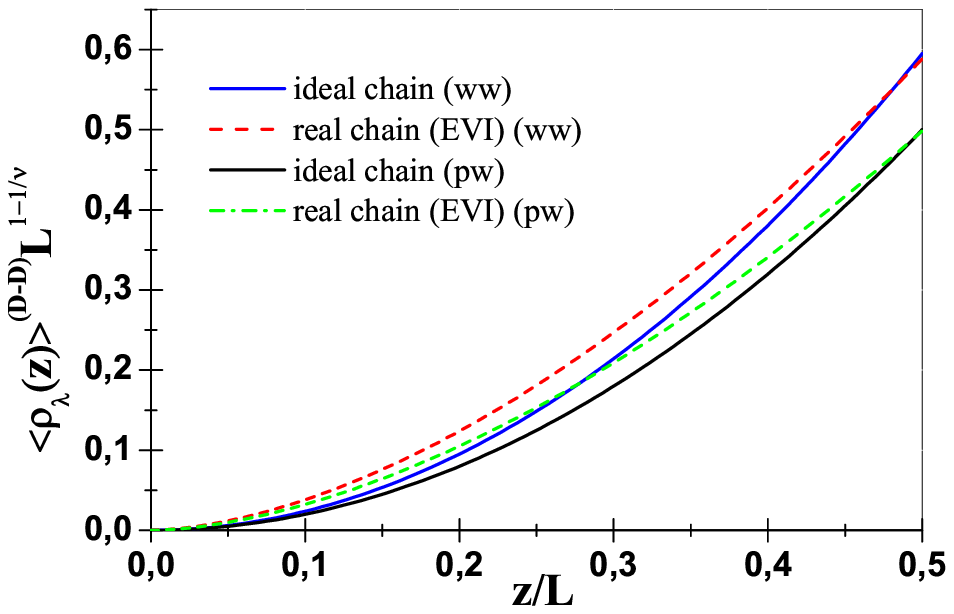}\hspace*{0.2cm}
\caption{The dimensionless value of the layer monomer density
$<\rho_{\lambda}(z)>L^{1-1/\nu}$ profiles for: ideal and real
polymer chains in the case of two repulsive walls (D-D b.c)(ww) with
$y=10$, and dilute polymer solution ($R_{x}=0.1\mu m$) of ideal and
real polymer chains in semi-infinite space containing repulsive
spherical particle of big radius $R=10.0 \mu m$ on the distance
$a=1.0 \mu m$ from the repulsive wall ($a=L$) (pw). The maximum is
at $L/2$. } \label{fig:1}
\end{center}
\end{figure}

\begin{figure}[ht!]
\begin{center}
\includegraphics[width=8.0cm]{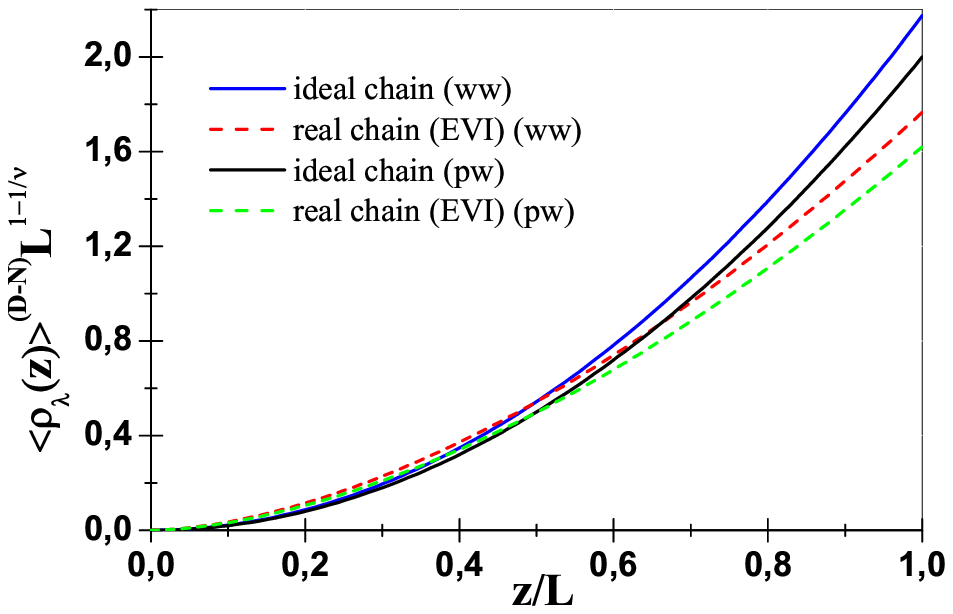}\hspace*{0.2cm}
\caption{The dimensionless value of the layer monomer density
$<\rho_{\lambda}(z)>L^{1-1/\nu}$ profiles for: ideal and real
polymer chains in the case of one repulsive one inert wall (D-N
b.c.) (ww) with $y=10$, and dilute polymer solution ($R_{x}=0.1\mu
m$) of ideal and real polymer chains in semi-infinite space
containing inert spherical particle of big radius $R=10.0 \mu m$ on
the distance $a=1.0 \mu m$ from the repulsive wall ($a=L$) (pw). The
maximum is near the distant inert wall. } \label{fig:2}
\end{center}
\end{figure}

In order to test the reliability of the obtained analytical results
it will be interesting to compare them with the recent results of
Monte Carlo calculations obtained by \cite{HG04} for the single
polymer chain trapped inside the slit of two repulsive walls.

In Ref. \cite{HG04} the lattice Monte Carlo algorithm on a regular
cubic lattice in $d=3$ dimensions, with $D$ lattice units in
$z$-direction and impenetrable boundaries was applied ($L=uD$ with
$u$ denoting the lattice spacing). The other directions obeyed
periodic boundary conditions. The correspondent reduced force in
accordance with (\cite{HG04}) can be written in the form:
\eq{\frac{f}{k_B\,T}\,= \,\frac{{\tilde{a}}
}{\nu\,\mu_{\infty}L}\,\sqrt{3}^{\frac{1}{\nu}}\,
\left(\frac{L}{R_x}\right)^{-\frac{1}{\nu}}, }  where parameter
${\tilde{a}}$ is a universal amplitude, $\mu_{\infty}$ is the
critical fugacity per monomer. Taking into account
Eq.(\ref{mn_dens_rel}) and the value $B_{ideal}$ from
(\ref{bideal}), the monomer density near the wall for ordinary RW is
scaled as: \be <\rho_{\lambda}(z)>_{G}=\frac{2 \pi^2 z^{2}}{L}
\left(\frac{L}{R_{x}}\right)^{-2} \label{mn_dens_mc_g},\ee where the
universal amplitude ${\tilde{a}}$ for the case of ideal chains was
found as ${\tilde{a}}\approx0.2741(2)$ (see Ref.\cite{HG04} ), which
is very close to the exact value, computed analytically in
\cite{E97} and equal to ${\tilde{a}}=\frac{\pi^2}{36}$. For the
ideal chain takes place: $\nu=0.5$, $\chi_d=1$ and
$\mu_{\infty}=\frac{1}{6}$. In Fig.3 this asymptotic behaviour for
narrow slits is clearly recovered by our results for ideal chains,
where the narrow slit limit is valid.
\begin{figure}[ht!]
\begin{center}
\includegraphics[width=8.0cm]{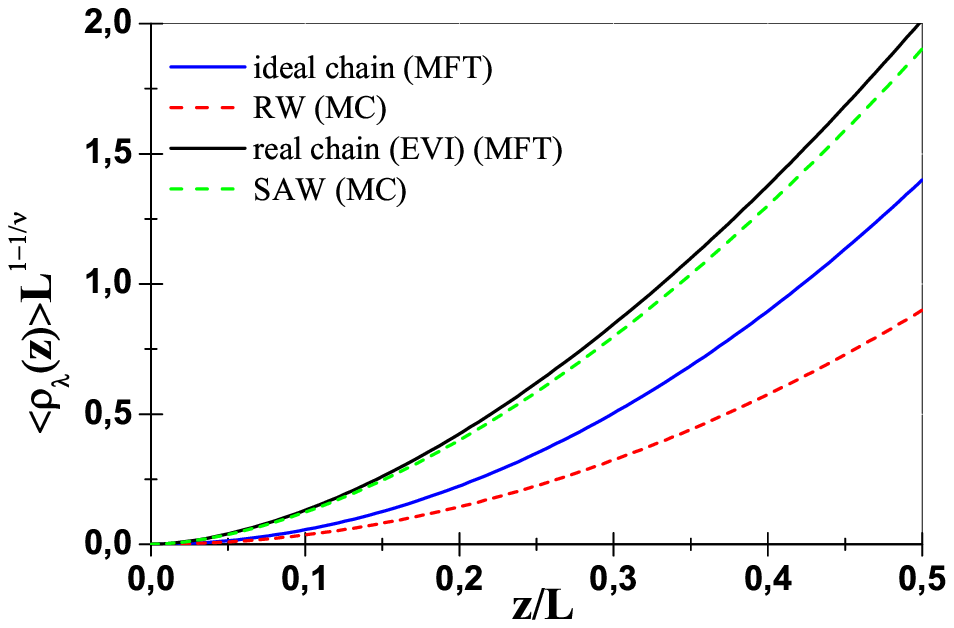}\hspace*{0.2cm}
\caption{Comparison of theoretical results with Monte Carlo
simulations for the layer monomer density $<\rho_{\lambda}(z)>$
profiles for: ideal chain and real polymer chain with EVI (MFT) in
the case of two repulsive walls (D-D b.c). RW(MC) and SAW(MC) are
the results obtained in Ref.\cite{HG04} for random walks and self
avoiding walks. The comparison of results are performed for the
case: $D=80$, $N=3500$ and $y\approx 2.34$ (RW); $D=120$, $N=3500$
and $y\approx 1.73$ (SAW) .} \label{fig:3}
\end{center}
\end{figure}

In the case of SAW in Ref. \cite{HG04} the value
${\tilde{a}}\approx0.448\pm 0.005$ was obtained. Taking into account
the values for $\nu\approx0.588$, $\chi_3^2\approx0.960$ (e.g.
\cite{CJ90}) and $\mu_{\infty}\approx0.2135$, $B_{real}= 1.70\pm
0.08$ (see Ref. \cite{HG04}), the correspondent monomer density for
SAW can be written as \be
<\rho_{\lambda}(z)>_{EVI}\,\approx\,\frac{15.44\,z^{1.7}}{L}\left(\frac{L}{R_x}\right)^{-1.7}\,\,.\label{mn_dens_mc_evi}
\ee  The result Eq.(\ref{mn_dens_mc_evi}) is presented in Fig.3 and
compared to our theoretical results for a trapped chain with EVI,
which are valid for the wide slit regime $y\gtrsim 1$. In order to
compare the obtained analytical results with recent results of MC
simulation we performed extrapolation of the MC results for
${\tilde{a}}$ and $\mu_{\infty}$ which are valid for the narrow slit
regime to the region of $y\gtrsim 1$. As it possible to see from
Fig.3, the results Eq.(\ref{mn_dens_mc_g}) and
Eq.(\ref{mn_dens_mc_evi}) correspond very well to our theoretical
predictions in the wide slit limit. One of the possible reasons for
remaining deviations with the results of Ref. \cite{HG04} is that
the chain in the MC simulation is too short in order to compare with
the results of the field-theoretical RG group approach.
Unfortunately, at the moment no simulations concerning one repulsive
and one inert wall exist.

\section*{Acknowledgments}
We gratefully acknowledge fruitful discussions with E.Eisenriegler.
 This work in part was supported by grant from the Alexander von Humboldt Foundation.

\end{document}